\documentclass[twocolumn,aps,showpacs,amsmath,amssymb]{revtex4}
\usepackage{graphicx}
\usepackage{dcolumn}
\usepackage{bm}

\begin{document}

\title{A stochastic model for heart rate fluctuations}

\author{Tom Kuusela}
  \email{tom.kuusela@utu.fi}
\author{Tony Shepherd}
\author{Jarmo Hietarinta}

\affiliation{Department of Physics, University of Turku, 20014
Turku, Finland}

\date{\today}

\begin{abstract}
Normal human heart rate shows complex fluctuations in time, which
is natural, since heart rate is controlled by a large number of
different feedback control loops. These unpredictable fluctuations
have been shown to display fractal dynamics, long-term
correlations, and 1/f noise. These characterizations are
statistical and they have been widely studied and used, but much
less is known about the detailed time evolution (dynamics) of the
heart rate control mechanism. Here we show that a simple
one-dimensional Langevin-type stochastic difference equation can
accurately model the heart rate fluctuations in a time scale from
minutes to hours. The model consists of a deterministic nonlinear
part and a stochastic part typical to Gaussian noise, and both
parts can be directly determined from the measured heart rate
data. Studies of 27 healthy subjects reveal that in most cases the
deterministic part has a form typically seen in bistable systems:
there are two stable fixed points and one unstable
one.
\end{abstract}


\maketitle

\section{Introduction}
Various methods and models have been used in attempts to
characterize the dynamics of the heart rate control mechanism. For
short time periods and under stationary conditions there are
successful models of heart rate and blood pressure regulation
\cite{Seidel,Voorde}, but the characterization of long-term
behavior has been a very difficult problem. Some models have been
introduced in order to explain long-term fluctuations, but usually
they can only describe well-controlled in vitro experiments, or
the models depend on large number of parameters, which cannot be
easily determined from experimental data \cite{Glass88}.
Furthermore, these models can predict only global statistical
features like scaling properties of power spectrum and
correlations \cite{Ivanov} and tell us very little about the
details of the time evolution.

Many features can be extracted from long time series of heart rate
measurements, quantities like entropy measures
\cite{Bettermann,Kaspar,Pincus95a,Pincus95b,Rezek,Richman,Zhang}
correlation dimension
\cite{Farmer,Fell,Grassberger,Kantz,Mayer,Yum}, detrended
fluctuations \cite{Peng93,Peng95,Iyengar}, fractal dimensions
\cite{Bassingthwaighte,Chau,Gough,Zhang}, spectrum power-law
exponents \cite{Bigger,Iyengar}) and symbolic dynamics complexity
\cite{Palazzolo,Voss95,Voss96}, but these are all purely
statistical characterizations and as such cannot provide us a
mathematical model of heart rate dynamics, not even a simple one.
However, some of these statistical methods do characterize the
complexity of the dynamics underlying the time series
\cite{Kuusela}, or are directly related to their fractal or
chaotic features. Mathematical analysis of many physiological
rhythms, including long-term heart rate fluctuations, has revealed
that they are generated by processes which must be nonlinear,
since linear systems can not produce such a complex behaviour
\cite{Glass01}. Nonlinear purely deterministic models can display
chaotic dynamics and generate apparently unpredictable
oscillations, but in practice it has not yet been possible to
extract such models from real noisy experimental data. It is also
possible that the underlying system is stochastic, i.e., the time
evolution of the system is subject to a noise source. [This kind
of dynamical noise is different from measurement noise, which is
mostly generated in the experimental apparatus.] In any case,
there is increasing evidence that noise, originated either from
the system itself or as a reflection of external influences, is
actually an integral part of the dynamics of biological systems
\cite{Collins,Hidaka,Mar}.

\begin{figure}
\includegraphics{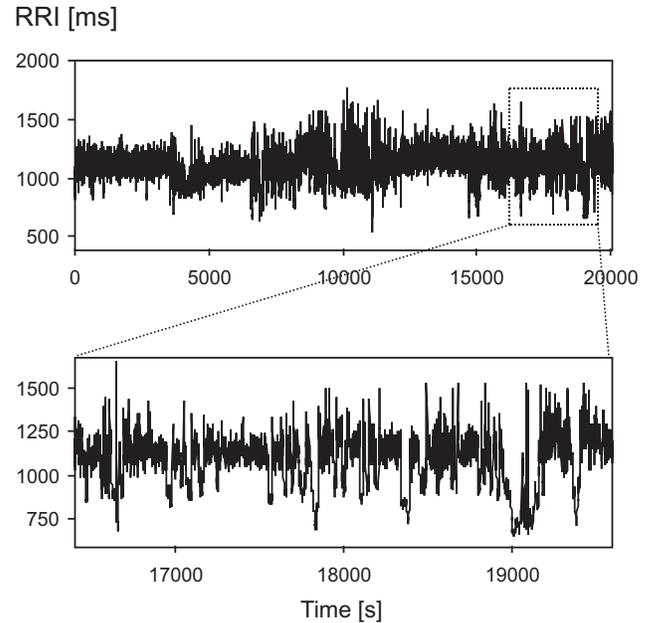}
\caption{\label{fig1} Typical R-R interval time series recorded at
night}
\end{figure}

\begin{figure*}
\includegraphics{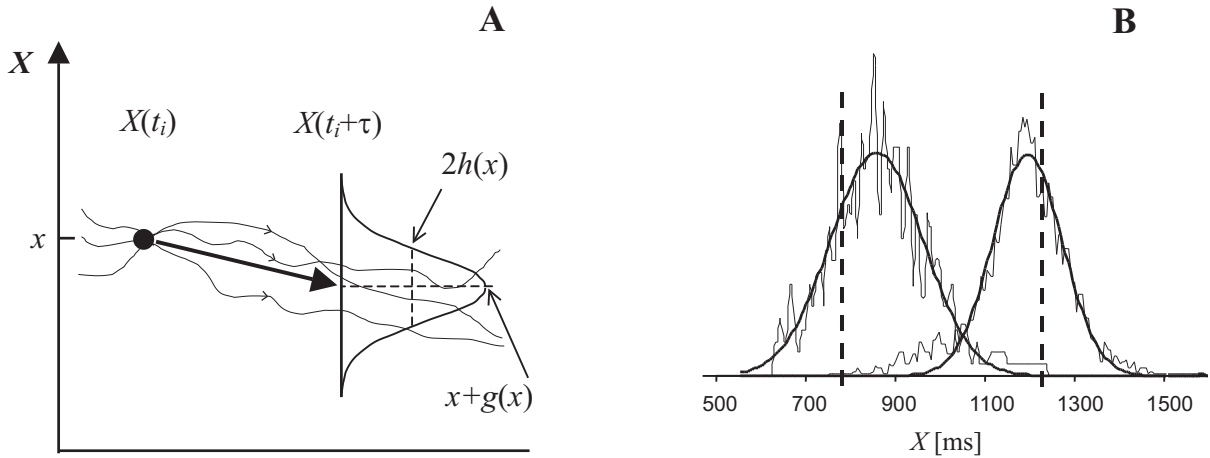}
\caption{\label{fig2}Schematic presentation of the method for
analyzing stochastic time series and calculating the deterministic
and stochastic part of the dynamics (Fig. A). Whenever the
trajectory of the system passes near certain point $x$ in the
state phase, i.e. $X(t_i) \approx x$, the future value $X(t_i +
\tau)$ of the trajectory is recorded. The distribution of these
values is fitted by a Gaussian function with the mean $x + g(x)$
and deviation $h(x)$, cf. equation (\ref{eq2}). This is repeated
for all $x$-values. On the right are two typical examples of the
distribution of future values, the initial $x$ values are marked
with dashed vertical bars and the fitted Gaussian curves with a
thick line (Fig. B).}
\end{figure*}

A typical R-R interval recording is shown in Fig.\ \ref{fig1}. The
time series is generated by recording a 24-hour electrocardiogram
and detecting the R-peak from each heartbeat, the R-R interval is
the time difference between two consecutive R-peaks. In the upper
panel of Fig.\ \ref{fig1} we have the R-R interval time series for
6 hours. We can see sections where the oscillations are rather
regular but there are also abrupt changes. In the lower panel of
Fig.\ \ref{fig1} we have zoomed into a part of the time series,
about $50$ minutes, and also on this time range we can see
apparently random oscillations with rapid changes.

It is well known that most short-time fluctuations of heart-rate
are generated by respiration (periods typically in the couple of
seconds range) and blood pressure regulation (so called Meyer
waves with periods of about 10 seconds \cite{Guyton}). In the
following we are not interested in these fast rhythms (which can
be analyzed quite well using linear or semi-linear models) but
rather in time scales from minutes to hours. We will show that in
this time range the dynamics of the heart-rate fluctuations can be
well described by a one-dimensional Langevin type difference
equation. This equation contains a deterministic part and additive
Gaussian noise, and we have found that it works well when the
delay parameter in the equation is in the range of 2--20 minutes.

\section{The Model}
An important and wide class of dynamic systems can be described by
the Langevin differential equation \cite{Kampen,Risken}
\begin{equation}
\dfrac{dX(t)}{dt}=g(X(t),t)+h(X(t),t)\Gamma(t) . \label{eq1}
\end{equation}
Here $X(t)$ represent the state of the system at time $t$,  the
function $g$ gives the nonlinear deterministic change, and the
last term $h$ is the amplitude of the stochastic contribution and
$\Gamma(t)$ stands for uncorrelated white noise with vanishing
mean. These kinds of stochastic differential equations always need
an interpretation rule for the noise term, normally one uses the
Ito interpretation \cite{Ito}. In general the functions $g$ and
$h$ could depend explicitly on time $t$. The equation (\ref{eq1})
can be easily generalized to higher dimensions. We will now show
that long-term behaviour of heart rate can be modeled using a {\it
difference} version of the Langevin equation \cite{Kampen}
\begin{equation}
X(t+\tau)=X(t)+g(X(t);\tau)+h(X(t);\tau)\Gamma(t) . \label{eq2}
\end{equation}
Here $X(t)$ again represents the state of the system, in this case
the R-R interval, at time $t$, and $\tau$ is the time delay. If
arbitrary small delays $\tau$ are possible then one can take the
limit $\tau\rightarrow 0$ and get the differential equation
(\ref{eq1}) [if the $\tau$ dependence is given by
$g(X(t);\tau)\approx \tau g(X(t))$], but in the present case it
will turn out that there is a minimum $\tau$ for which the model
(\ref{eq2}) seems to be valid. We assume that $g$ and $h$ do not
have explicit time dependence but they may depend on the delay $
\tau$. It is convenient to extract the term $X(t)$ in the
deterministic part, as is done in (\ref{eq2}), then a nonzero
$g(X(t);\tau)$ stands for changes in the state of the system. An
essential feature of models of the above type is that for time
evolution we only need to know the state at one given moment and
not its evolution in the past, i.e., they are Markovian
\cite{Hanggi,Kampen}.

The computational problem is now to determine the functions $g$
and $h$ from measured time series and to verify that the
description using (\ref{eq2}) is accurate. The principle of the
method is very simple \cite{Gradisek,Siegert}: at every time $t_i$
when the trajectory of the system meets an arbitrary but fixed
point $x$ in state space, we look at the future state of the
system at time $t_ i +\tau$. The set of these future values (for a
chosen $x$ and $\tau$) has a distribution in the state space and
from this distribution we can determine the deterministic part
$g(x)$ and the stochastic part $h(x)$, see Fig.\ \ref{fig2}A. In
practice we first divide the range of the dynamical variable $X$
into equal boxes. By scanning the whole measured time series we
check when $X$ is inside a given box $x$, i.e., $\vert
X(t_i)-x\vert\leq\Delta x$ , where $x$ is the middle value of the
box and $\Delta x$ is the half width of the box. When $X$ is found
on the box, we look at the future value of the variable,
$X(t_i+\tau)$, where $\tau$ is the fixed delay parameter. Since
the trajectory of the system passes each box several times, we can
calculate the distribution of the future values $X(t_i+\tau)$ for
each box $x$. If we assume that the noise is Gaussian, we can fit
a Gaussian function on each distribution, and as a result we get
the mean and the deviation parameters for each $x$; the mean of
this distribution is equal to $x+g(x)$ and the deviation is equal
to $h(x)$ \cite{Friedrich,Timmer}. A typical case is given in
Fig.\ \ref{fig2}B, and it shows that the distribution is actually
very well described by Gaussian noise (the correlation is better
than 0.95; the correlation is calculated as
$\sqrt{1-S_{res}/{S_{tot}}}$, where $S_{res}$ is the sum of the
squared residuals and $S_{tot}$ is the variance). From the given
data we can in this way determine the functions $g(X)$ and $h(X)$
needed in the stochastic model (\ref{eq2}). It should be noted
that we can calculate only the absolute value of $h(X)$ since the
deviation parameter found from the fitted Gaussian function is in
squared form.

In our analysis we have used R-R interval time series of
$22$--$24$ hours, corresponding to $80.000$--$100.000$ data
points. Our data is actually interval data, i.e., it consists of a
sequence of R-R interval values. It is then convenient to count
the delay in our analysis in terms of heart-beats rather than
seconds, i.e., we have not used cumulative time as time variable
but the beat index. However, since the R-R interval values vary a
lot within the used delay range, the beat index actually gives a
delay as if computed with the average beat-rate.  We have tested
both methods and found only minor differences between them (in the
details of the functions $g$ and $h$). We will show later that the
functional forms of $g$ and $h$ are quite insensitive on the time
delay, and since this holds for both methods we will use the more
convenient beat index.

\begin{figure*}
\includegraphics{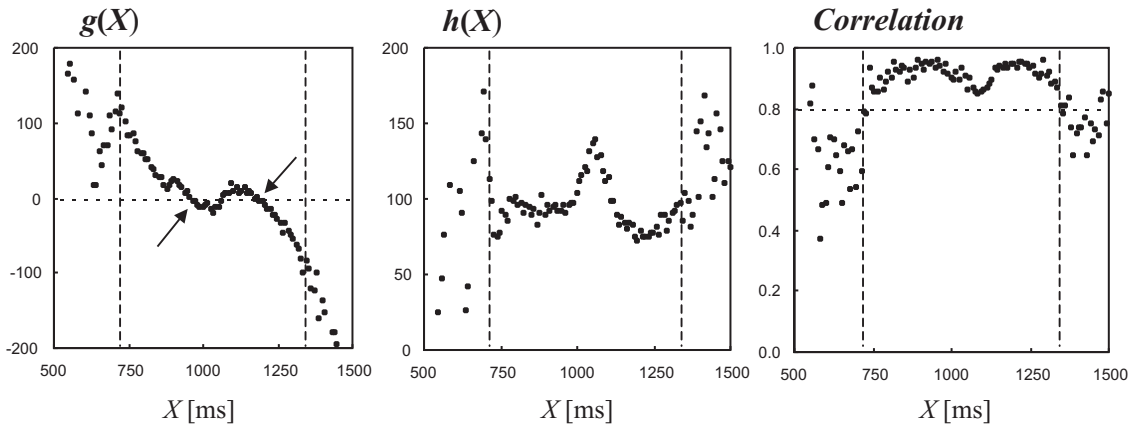}
\caption{\label{fig3}Typical results derived from R-R interval
time series using time delay $\tau = 500$. We have shown the
deterministic part $g(X$) (the left panel), the stochastic part
$h(X)$ (the middle panel) and the correlation coefficient of the
distribution (the right panel) as a function of the dynamical
variable $X$. The range corresponding to the correlation threshold
level of $0.8$ is marked with the vertical lines.}
\end{figure*}

\section{Results}
\subsection{A typical case}
In Fig.\ \ref{fig3} we have presented results obtained for a
particular case using the method described earlier. The value of
the delay parameter $\tau$ was $500$ beats, and the number of
boxes used to construct local distributions was $150$.
Distributions were fitted using Gaussian function. The $g(X$)
function, the deterministic part of the system, is displayed on
the left panel in Fig 3. It has a very clear and simple functional
form (between the vertical lines) which is typical for systems
exhibiting bistable behaviour \cite{Berge,Kampen}. The function
crosses the zero line three times, these crossings are the fixed
points of the system. The fixed points marked with arrows are
stable: without any noise term these points attract all nearby
states, because the control function $g(X)$ is locally decreasing.
The middle fixed point is repulsive. Due to the stochastic part
the system has a tendency to jump between the stable points if the
amplitude of the noise is high enough. Far away from the stable
points $g(X)$ increases or decreases strongly forcing the system
rapidly back to oscillate around the stable points. The amplitude
of the stochastic part of the system, function $h(X)$, is almost
constant except between the stable points where it has a clear
maximum (the middle panel in Fig.\ \ref{fig3}).  One
interpretation is that the system has a larger inherent freedom to
oscillate randomly when the trajectory is between the stable
points but outside this range the character of the system is more
deterministic. From the physiological point of view this kind of
dynamics can be useful since it lets the R-R interval to wander
most of the time but prevents it from escaping too far away from
the normal range. On the right panel in Fig.\ \ref{fig3} we have
shown the correlation coefficient of each local distribution. Most
of the time the correlation is remarkably high, about
$0.85$--$0.95$, but near the largest and smallest $X$ values there
are only rather few data points and therefore the corresponding
distributions do not have clear Gaussian shape resulting with
lower correlation.  The high average correlation value is a clear
indication that the noise in this system is really Gaussian type.
We have used the value of $0.8$ as a threshold level, and the
corresponding range is marked with the vertical lines in Fig.\
\ref{fig3}.

\begin{figure*}
\includegraphics{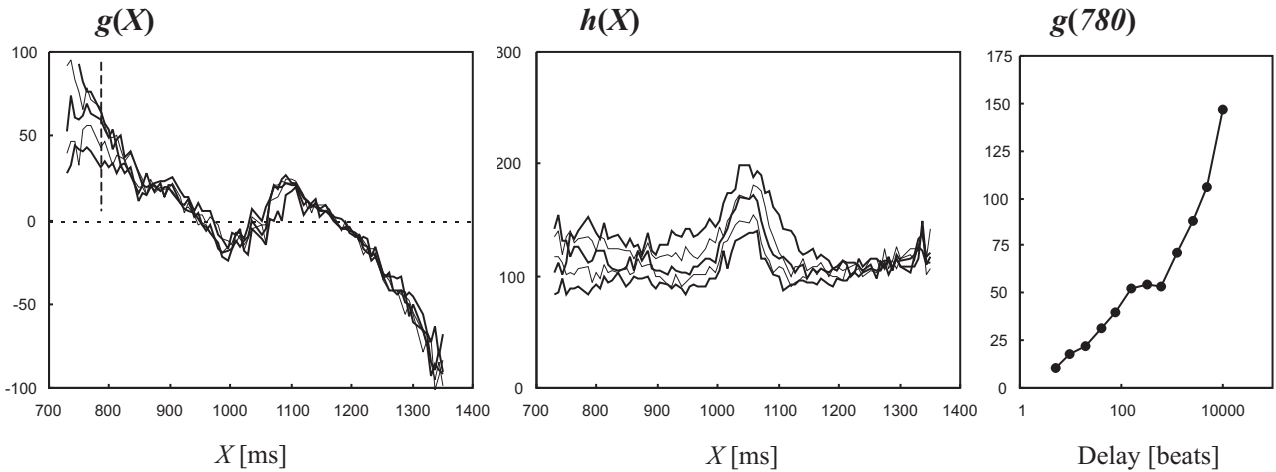}
\caption{\label{fig4}Examples of the deterministic part $g(X$)
(the left panel) and the stochastic part $h(X$) (middle panel)
calculated with various values of the delay parameter $\tau$ :
$40$ (thick line), $80$ (thin line), $160$ (thick line), $320$
(thin line) and $640$ (thick line). The values of the $g(X$)
function at $X = 780$ ms (marked with vertical dashed line in the
left panel) are plotted as a function of the delay in the right
panel, there is a plateau around a delay of $100$--$1000$ beats.}
\end{figure*}

What is remarkable in this description is that the functional
forms of $g(X)$ and $h(X)$ are fairly independent of the delay
parameter $\tau$ in a rather extensive delay range, typically
$100$--$1000$ beats (corresponding to $2$--$20$ minutes). In Fig.\
\ref{fig4} we have plotted the functions $g(X)$ and $h(X)$ for a
range of $\tau$ values. The $g$-function is practically $\tau$
independent, except for the shortest R-R intervals, where some
cumulative effects show up. The $h$-function seems to grow very
slowly as $\tau$ increases. For still smaller delay values $g(X)$
is more flat and $h(X)$ is more scattered, and for longer delays
$g(X)$ is typically a straight line and $h(X)$ is constant.
Behavior at these extremes can be easily understood by recalling
that when the time scale is small, the heart rate system is
clearly multidimensional depending directly on blood pressure,
respiration and other rapidly changing physiological variables and
our 1-dimensional description is no longer valid. On the other
hand, if the delay parameter is very large, we cannot reconstruct
the local dynamics in terms of local distributions, we just get
the global distribution that is independent of dynamics and no
longer Gaussian. In the right panel of Fig.\ \ref{fig4} we have
given the values of the $g(X)$ function at $X = 780$ ms (marked
with a vertical dashed line in the left panel) computed with
delays of $5$--$10240$ beats. We can see a plateau in the delay
range of $100$--$1000$ beats which means that the $g(X)$ curves
for these delays are bundled. In principle the curves for a delay
of $2\tau$ should be obtainable by iterating (\ref{eq2}) with
delay $\tau$. Direct numerical calculations of joint probabilities
using experimentally determined $g(X)$ (within $100$--$1000$ beats
delay range) indicate that $g(X)$  and $h(X)$ do not change
significantly in one iteration, mostly because in our case the
Gaussian distribution is not so narrow. In general iterations tend
to sharpen the bends in $g(X)$ and this feature is indeed visible
in Fig.\ \ref{fig4}. The small $\tau$-dependency of $g$ and $h$ in
the range of short R-R intervals can then be interpreted either as
the expected result from repeated iterations, or as a sign of
higher order dynamics: possibly the heart rate regulation system
is more complex when the system must readjust at a fast heart
rate.

\begin{figure*}
\includegraphics{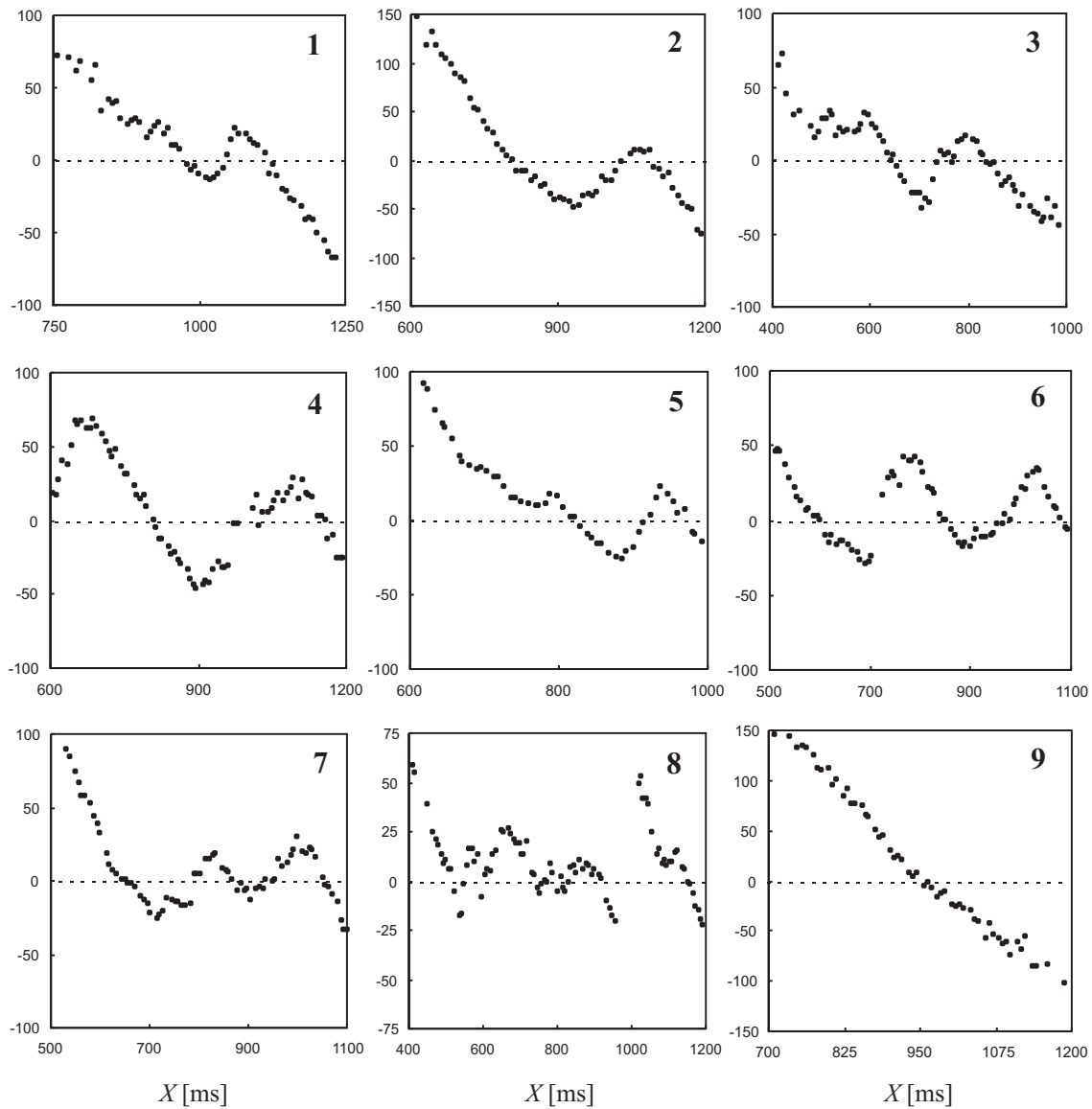}
\caption{\label{fig5}Typical deterministic functions $g(X)$
derived from different subjects. The cases 1 - 5 represent simple
bistable situation, the cases 6 and 7 have three stable points,
the case 8 is multistable, and the case 9 has only a single stable
fixed point.}
\end{figure*}

\subsection{Variation between subjects}
In order to find whether different subjects have are any
common features in the deterministic and stochastic parts $g(X)$
and $h(X)$ we analyzed the data from $27$ healthy subjects of
various age and gender [$18$ cases from PhysioBank
\cite{Goldberger} and $9$ cases from Kuopio University Hospital].
Analyses were done using the same parameter values as in Fig.\
\ref{fig3}. The deterministic part, the $g(X)$ function, is
displayed in Fig.\ \ref{fig5} for a set of $9$ typical cases. The
most common form for this function is the bistable type, already
shown in Fig.\ \ref{fig3}, where the $g(X)$ function has three
zeroes, and $60$\% of all cases can be classified to this group
(cases 1--5 in Fig.\ \ref{fig5}). The next most common group,
$25$\% of all cases, has a $g(X)$ function with $5$ zeroes, a kind
of double pitchfork system (cases 6 and 7 in Fig.\ \ref{fig5}). We
also found $3$ cases where the $g(X)$ function seems to have even
more zeroes (case 8 in Fig.\ \ref{fig5}). Only very few cases
could not be clearly classified as bi- or multistable. In these
cases it can be difficult to interpret the results. It is possible
that the dynamical variable did not explore the whole state phase,
and therefore we can see only part of the $g(X)$ function; for
example case 9 in Fig.\ \ref{fig5}, where the system has only one
stable fixed point and no unstable points at all, can be  an
example of this. The stochastic parts (function $h(X)$) are fairly
similar: they are almost constant except that in all cases there
are maxima on the R-R interval ranges between the stable fixed
points of the deterministic part, as in the example in Fig.\
\ref{fig3}.

\begin{figure*}
\includegraphics{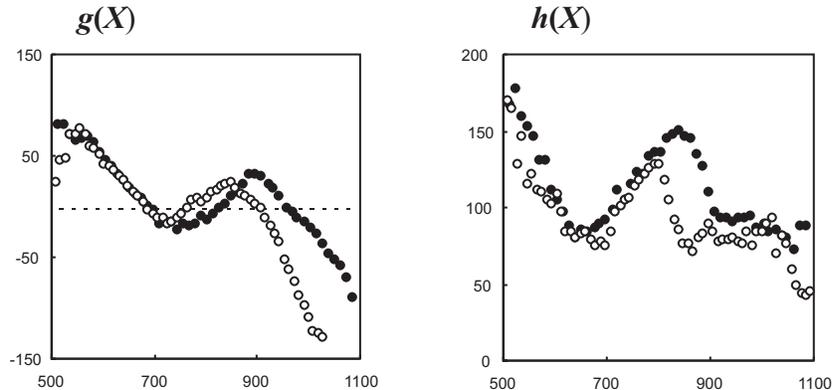}
\caption{\label{fig6}The deterministic parts $g(X)$ (left panel)
and stochastic parts $h(X)$ (right panel) computed from the R-R
interval time series recorded from the same subject on different
days. The data from the first recording is marked with solid dots
and data from the second recording, 4 days later, with open dots.}
\end{figure*}

The description given by equation (\ref{eq2}) contains both a
deterministic and a stochastic component. It is an important to
realize that the stochastic part is not a small perturbation but
in fact forms an essential part of the description, furthermore it
is $10$--$20$ times higher than the measurement noise (uncertainty
in detecting the position of the R-peak), which is typically only
$2$--$5$ ms. One way to compare the deterministic and stochastic
components is to note that the size of the bend in the $g(X)$
function is of the order of $30$ to $50$ ms, while the average
size of the $h(X)$ function is about $70$ to $110$ ms, as can be
seen in Fig.\ \ref{fig3}. [The extraction of small details in the
$g(X)$ function under such noise is of course possible only
because the noise is so cleanly Gaussian.] On the other hand, the
distance between the stable fixed points in the $g(X)$ function is
of the order of $50$ to $250$ ms, and therefore the probability
that the systems jumps between stable points is not extremely
high, but nevertheless possible. It is also possible that external
factors drive the system from one stable point to another, since
during night-time the mean R-R interval is typically longer than
during day-time [although the R-R interval can abruptly jump to
the faster rate also during the night, as can be seen on the lower
panel in Fig.\ \ref{fig1}].

\subsection{Same subject at different times}
If the model (\ref{eq2}) were to describe true heart rate
dynamics, the functions $g(X)$ and $h(X)$ should have some
constant features specific for each subject. In order to look at
this aspect we made two recordings from the same subject within 4
days, the results are shown in Fig.\ \ref{fig6}. In general the
deterministic and stochastic parts from different recordings are
remarkably similar both having clear bistable character. In the
R-R interval range of $500$--$800$ ms the results are almost
identical and the only difference seems to be a scaling towards
the shorter R-R intervals in the $800$--$1100$ ms range of the
second recording. In the first recording the mean value of R-R
interval calculated over the 24 hour period was $781$ ms and in
the second one $726$ ms. Therefore in the second recording the
shortest R-R intervals are significantly more frequent and this
can affect on the analysis results. These deviations could also
reflect true changes on the underlying control system: it is well
know that there are daily variations on functions of the autonomic
nervous system.

\begin{figure*}
\includegraphics{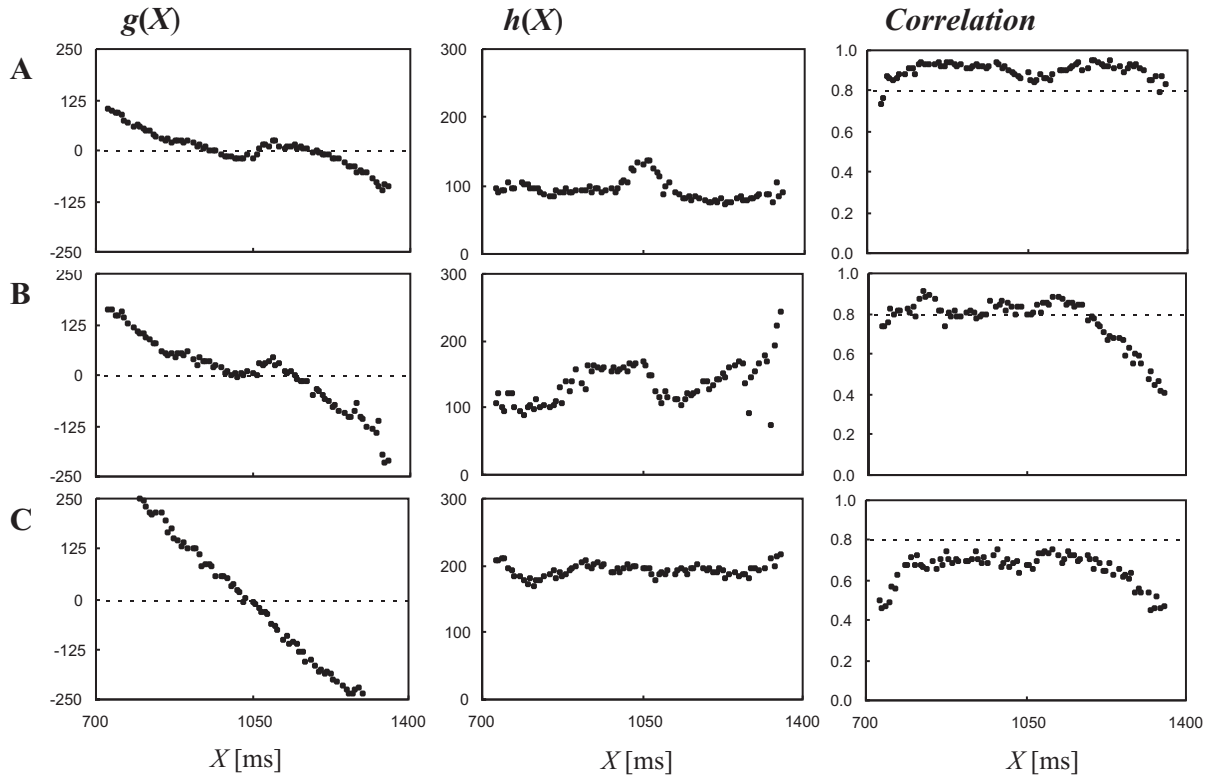}
\caption{\label{fig7}The deterministic part $g(X)$ (left column),
stochastic part $h(X)$ (middle column) and correlation coefficient
(right column) for the original data (row A) and for two surrogate
versions (B and C). For surrogate data the original data has been
shuffled using section sizes of $800$ (B) and $400$ (C) data
points.}
\end{figure*}

\subsection{Surrogate analysis}
As a further validity check we also performed surrogate analysis
\cite{Schreiber,Theiler} in order to eliminate the possibility
that the results are generated just from peculiar distribution of
the R-R intervals imitating real dynamics. For this purpose the
data was shuffled by dividing it into sections of equal size which
were then repositioned randomly. As a result we get a new time
series where the dynamical structure has been partially destroyed
depending on the section size. Results of this surrogate analysis
are shown in Fig.\ \ref{fig7}. The top panels display the
deterministic $g(X)$ and stochastic $h(X)$ parts of the system and
the correlation coefficient without any data shuffling (row A in
Fig.\ \ref{fig7}). On the next row (row B in Fig.\ \ref{fig7}) we
have used sections of $800$ data points for shuffling. There are
only small changes in the deterministic part, but the correlation
has decreased noticeably. When the section size is $400$ (row C in
Fig.\ \ref{fig7}) we can no longer see the bistable character in
the deterministic part, the stochastic part is flat with higher
mean level, and the average level of the correlation coefficient
has dropped well below our threshold value $0.8$. With still
smaller section sizes the results do not change any further. In
this analysis we have used the same delay of $500$ data points as
previously and when the section size used in the shuffling process
is less than this delay all dynamical properties disappear, as
expected in the case of true time evolution. Therefore we conclude
that our results are derived from the dynamical properties of the
heart beat data, and not from their overall statistical
characteristics.

\section{Conclusion}
Our results indicate that the human heart-rate control dynamics
can be accurately modeled with the 1-dimensional stochastic
difference equation (\ref{eq2}), where the time delay parameter is
within $2$--$20$ minutes. Stochasticity is an integral part of the
dynamics, and in this delay range the effects of other variables
are either embedded into the stochastic part of the system or
averaged over time with no net effect. It is remarkable that the
form of the control function $g(X)$ is similar from case to case.
Their typically bistable character is also well justified on
common physiological grounds. From this initial study we cannot
yet identify what kind of dynamical structure is typical for
healthy subjects (although our results already indicate that a
simple bistable system is most common feature) and therefore the
model cannot yet be used directly for clinical work, for that
purpose one needs extensive demographic studies. We can
nevertheless speculate that the form of the control function
$g(X)$ should tell us something about the health of the subject.
Also, some of the current knowledge based on statistical measures
of heart rate time series can probably be explained within the
framework of our model. Another interesting observation is the
importance of the stochastic part, it could be the result of
integrating the effects of a more detailed control mechanism over
time, but it could also reflect some truly stochastic internal and
external influences.

\begin{acknowledgments}
We thank T. Laitinen from Kuopio University Hospital, Department
of Clinical Physiology, for providing nine electrocardiogram
recordings. This work was partially supported by the Academy of
Finland.
\end{acknowledgments}

\bibliography{Stochastic}

\end{document}